\documentclass[12pt,a4]{article}
\usepackage{amsmath}
\usepackage{bm}
\usepackage{amsfonts}
\usepackage{amssymb}
\usepackage{graphics}
\usepackage[normal]{caption2}
\usepackage{subfigure}
\usepackage{rotating}
\usepackage{citesort}
\def\be{\begin{equation}}
\def\ee{\end{equation}}
\def\ba{\begin{array}}
\def\ea{\end{array}}
\def\bea{\begin{eqnarray}}
\def\eea{\end{eqnarray}}

\begin{document}
\baselineskip 20pt \setlength\tabcolsep{2.5mm}
\renewcommand\arraystretch{1.5}
\setlength{\abovecaptionskip}{0.1cm}
\setlength{\belowcaptionskip}{0.5cm}
%%%%%%%%%%%%%%%%%%%%%%%%%%%%%%%%%%%%%%%
\pagestyle{empty}
\newpage
\pagestyle{plain} \setcounter{page}{1} \setcounter{lofdepth}{2}
\begin{center} {\large\bf Transverse in-plane flow: a new probe of symmetry energy in Fermi energy region}\\
\vspace*{0.4cm}

{\bf Sakshi Gautam} and {\bf Rajeev K. Puri}\footnote{Email:~rkpuri@pu.ac.in}\\
{\it  Department of Physics, Panjab University, Chandigarh -160
014, India.\\}
\end{center}

\section*{Introduction}
After about three decades of extensive efforts in both nuclear
experiments and theoretical calculations, the equation of state
(EOS) of isospin symmetric matter is well understood through
experiments of collective flow  and subthreshold kaon production.
Nowadays, the nuclear EOS of asymmetric nuclear matter has
attracted a lot of attention. The EOS of asymmetric nuclear matter
can be described approximately by
\begin{equation}
E(\rho, \delta) = E_{0}(\rho,
\delta=0)+E_{\textrm{sym}}(\rho)\delta^{2}
\end{equation}
where
 $\delta$ = $\frac{\rho_{n}-\rho_{p}}{\rho_{n}+\rho_{p}}$ is isospin asymmetry,
  E$_{0}$($\rho$, $\delta$) is the energy of pure symmetric nuclear matter, and E$_{\textrm{sym}}$($\rho$)
  is the symmetry energy, where E$_{\textrm{sym}}$($\rho$$_{0}$) = 32 MeV is the symmetry energy at normal
  nuclear matter density. The symmetry energy is E($\rho$,1) - E$_{0}$($\rho$,0), i.e., the difference
  of the energy per nucleon between pure neutron matter and symmetric nuclear matter.
  The symmetry energy is important not only to the nuclear physics community
  as it sheds light on the structure of radioactive nuclei, reaction dynamics induced
by rare isotopes, but also to astrophysicists since it acts as a
probe for understanding the evolution of massive stars and the
supernova explosions. The existing and upcoming radioactive ion
beam (RIB) facilities lead a way in understanding nuclear symmetry
energy. Experimentally, symmetry energy is not a directly
measurable quantity and has to be extracted from observables which
are related to symmetry energy. Therefore, a crucial task is to
find such observables which can shed light on symmetry energy. A
large number of studies on the symmetry energy of nuclear matter
have been done during the past decade \cite{zhang05}. In this
paper, we aim to see the sensitivity of collective transverse
in-plane flow to symmetry energy and also to see the effect of
different density dependencies of symmetry energy on the same. The
various forms of symmetry energy used in present study are:
E$_{\textrm{sym}} \propto F_{1}(u)$, E$_{\textrm{sym}} \propto
F_{2}(u)$, and E$_{\textrm{sym}} \propto F_{3}(u)$, where \emph{u}
= $\frac{\rho}{\rho_{0}}$, F$_{1}(u) \propto u$, F$_{2}(u) \propto
u^{0.4}$, F$_{3}(u) \propto u^{2}$, and F$_{4}$ represents
calculations without symmetry energy. The study is carried out
with IQMD model \cite{hart98}.

\section*{Results and discussion}
We simulate several thousands events for the neutron-rich system
of  $^{48}\textrm{Ca}$+$^{48}\textrm{Ca}$ at energies of 100, 400,
and 800 MeV/nucleon at impact parameter of
b/b$_{\textrm{max}}$=0.2-0.4.
\par
In Fig. 1 we display the time evolution of
$<p_{x}^{\textrm{dir}}>$ for different symmetry energies used in
this paper at 100 MeV/nucleon for particles lying in the different
bins.  To understand the sensitivity of transverse momentum to the
symmetry energy as well as its density dependence in the Fermi
energy region, we calculate the transverse flow of particles
having $\frac{\rho}{\rho_{0}}$ $<$ 1 (bin 1) and particles having
$\frac{\rho}{\rho_{0}}$ $\geq$ 1 (bin 2). Dotted, solid and dashed
lines represent all particles, bin1 and bin2. Panels (a), (b), and
(c) are for E$_{\textrm{sym}} \propto \rho, \rho^{0.4}$, and
$\rho^{2}$, respectively. Panel (d) is for calculations without
symmetry energy. The total $<p_{x}^{\textrm{dir}}>$ is negative
during the initial stages and continues decreasing till 30 fm/c
which indicates dominance of attractive interaction. In panels (a)
and (b), it becomes positive whereas in panels (c) and (d) it
remains negative during the course of the reaction. If we look at
$<p_{x}^{\textrm{dir}}>$ of particles lying in bin 1 for F$_{1}
(u)$ [Fig. 1(a)] and F$_{2} (u)$ [Fig. 1(b)] in the time interval
0 to about 20-25 fm/c, we see that it remains positive. It
increases with time upto 15 fm/c and reaches a peak value. This is
because in the spectator region (where high rapidity particles
lie) the repulsive symmetry energy will accelerate the particles
away from the overlap zone in the transverse direction. After 15
fm/c, $<p_{x}^{\textrm{dir}}>$ (of particles in bin 1) begins to
decrease. This is because these particles will now be attracted
toward the central dense zone. From 10 to 20 fm/c particles in bin
2 continue increasing in the midrapidity region. In the case of
F$_{1} (u)$ and F$_{2} (u)$, particles which enter the central
dense zone (bin 2) have already a high positive value of
$<p_{x}^{\textrm{dir}}>$ (i.e., going away from the dense zone).
So, the attractive mean field has to decelerate the particles
first, make them stop, and then accelerate the particles back
toward the overlap zone. At about 20-25 fm/c particles from bin 1
have zero $<p_{x}^{\textrm{dir}}>$ [see shaded area in Fig. 1(a)
and 1(b)]. Up to 30 fm/c, particles feel the attractive mean field
potential after which the high density phase is over; i.e., in
case of F$_{1} (u)$ and F$_{2} (u)$ between 0 and 30 fm/c,
particles from bin 1 are accelerated toward the overlap zone only
for a short time interval of about 5 fm/c, whereas for the case of
F$_{3} (u)$ [Fig. 1(c)] and F$_{4}$ [Fig. 1(d)] between 0-30 fm/c,
particles from bin 1 are accelerated towards the overlap zone for
a longer time interval of about 20 fm/c between 10-30 fm/c.
Moreover, the $<p_{x}^{\textrm{dir}}>$ of particles lying in the
bin 1 [for F$_{3} (u)$ and F$_{4}$] follows a similar trend. This
is because for $\rho/\rho_{0} < $ 1 the strength of symmetry
energy F$_{3} (u)$ will be small and so there will be less effect
of symmetry energy on the particles, which is evident from Fig.
1(c) where one sees that the $<p_{x}^{\textrm{dir}}>$ remains
about zero during the initial stages between zero to about 10
fm/c. The $<p_{x}^{\textrm{dir}}>$ due to particles in bin 2
(dashed line) decreases in a very similar
 manner for all the four different symmetry energies between 0 and 10 fm/c. Between 10 and 25 fm/c, $<p_{x}^{\textrm{dir}}>$ for
 F$_{3} (u)$
 and F$_{4}$ decreases more sharply as compared to in
case of F$_{1} (u)$ and F$_{2} (u)$. This is because in this time
interval particles from bin 1 enter into bin 2. As discussed
earlier, $<p_{x}^{\textrm{dir}}>$ of particles entering bin 2 from
bin 1 in the case of F$_{1} (u)$ and F$_{2} (u)$ will be less
negative due to stronger effect of symmetry energy as compared to
the case of F$_{3} (u)$ and F$_{4}$ \cite{gaum1}.

\begin{figure}[!t] \centering
 \vskip 0cm
\includegraphics[width=12cm]{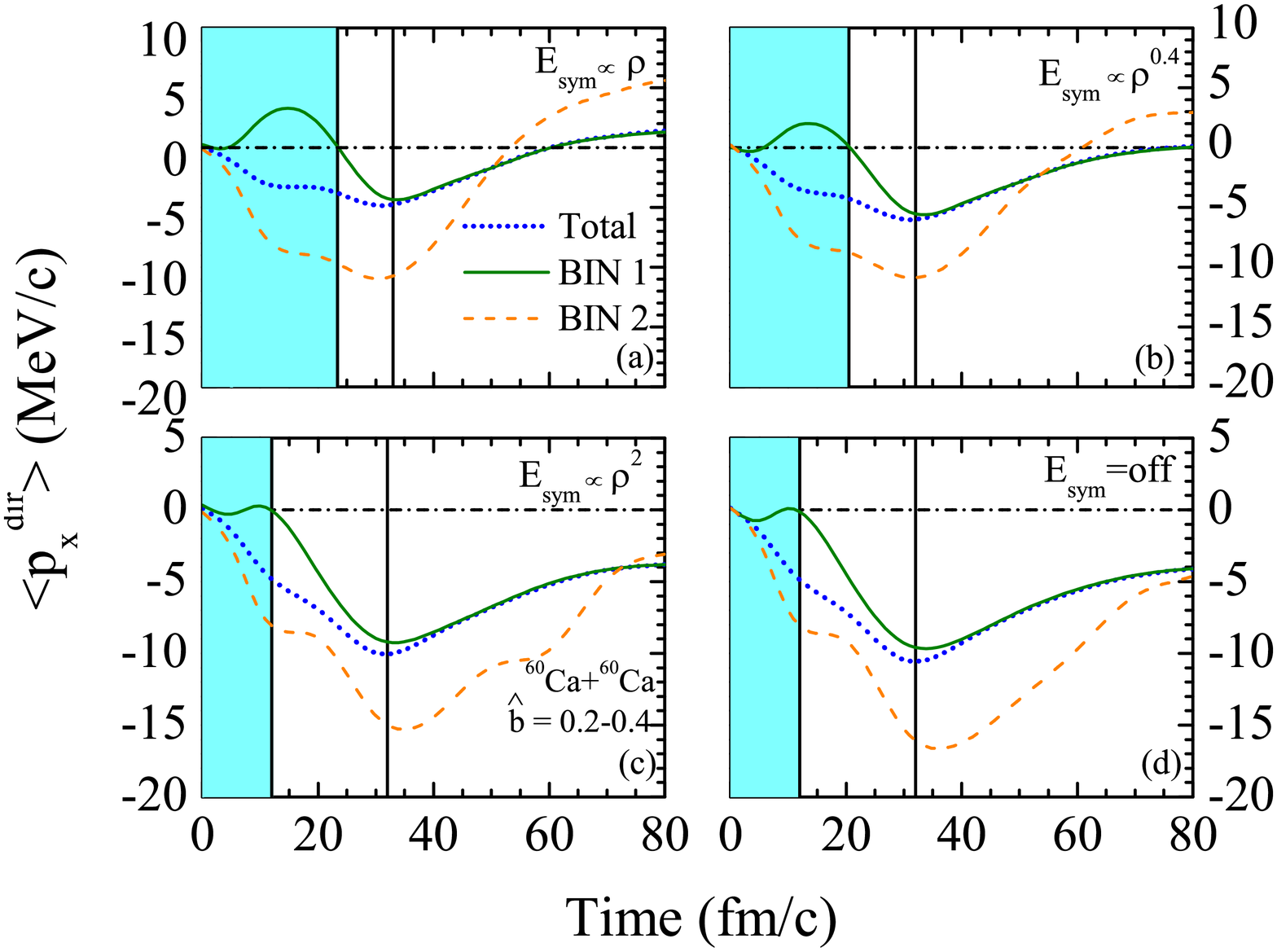}
 \vskip -0cm
\caption{The time evolution of $<p_{x}^{\textrm{dir}}>$ for
different forms of symmetry energy for different bins at
b/b$_{\textrm{max}}$=0.2-0.4.}\label{fig2}
\end{figure}

\section*{Acknowledgments}
 This work has been supported by a grant from Centre of Scientific
and Industrial Research (CSIR), Govt. of India.

\end{document}